\documentclass[useAMS,usenatbib,usegraphicx]{mn2e}

\title[Constraints on Planet Hosting Binary GJ\,86]{Orbital and Evolutionary Constraints on the Planet Hosting Binary GJ\,86 
from the {\em Hubble Space Telescope}}

\author[J. Farihi et al.]{J. Farihi$^{1,2}$\thanks{E-mail: jfarihi@ast.cam.ac.uk}\thanks{STFC Ernest Rutherford Fellow},
Howard E. Bond$^{3,4}$\thanks{Current address: 9615 Labrador Lane, Cockeysville, MD 21030, USA}, 
P. Dufour$^5$,
N. Haghighipour$^6$,
G. H. Schaefer$^7$,
\newauthor 
J. B. Holberg$^8$
M. A. Barstow$^2$,
M. R. Burleigh$^2$
\medskip\\
$^1$Institute of Astronomy, University of Cambridge, Cambridge CB3 0HA\\
$^2$Department of Physics \& Astronomy, University of Leicester, Leicester LE1 7RH\\
$^3$Space Telescope Science Institute, Baltimore, MD 21218, USA\\
$^4$Department of Astronomy \& Astrophysics, Pennsylvania State University, University Park, PA 16802, USA\\
$^5$D\'epartement de Physique, Universit\'e de Montr\'eal, Montr\'eal, QC H3C 3J7, Canada\\
$^6$Institute for Astronomy, University of Hawaii-Manoa, Honolulu, HI 96822, USA\\
$^7$CHARA Array of Georgia State University, Mount Wilson Observatory, CA 91023, USA\\
$^8$Lunar and Planetary Laboratory, University of Arizona, Tucson AZ 85721, USA}

\begin{document}

\date{}

\maketitle

\label{firstpage}

\begin{abstract}

This paper presents new observations of the planet-hosting, visual binary GJ\,86 (HR\,637) using the {\em Hubble Space 
Telescope}.  Ultraviolet and optical imaging with WFC3 confirms the stellar companion is a degenerate star and indicates 
the binary semimajor axis is larger than previous estimates, with $a\ga28$\,AU.  Optical STIS spectroscopy of the secondary
reveals a helium-rich white dwarf with C$_2$ absorption bands and  $T_{\rm eff}=8180$\,K, thus making the binary system 
rather similar to Procyon.  Based on the 10.8\,pc distance, the companion has 0.59\,$M_{\odot}$ and descended from a 
main-sequence A star of 1.9\,$M_{\odot}$ with an original orbital separation $a\ga14$\,AU.  If the giant planet is coplanar 
with the binary, the mass of GJ\,86Ab is between 4.4 and 4.7 $M_{\rm Jup}$.

The similarity of GJ\,86 and Procyon prompted a re-analysis of the white dwarf in the latter system, with the tentative conclusion 
that Procyon hosts a planetesimal population.  The periastron distance in Procyon is 20\% smaller than in $\alpha$\,Cen\,AB, 
but the metal-enriched atmosphere of Procyon\,B indicates that the planet formation process minimally attained 25\,km bodies,
if not small planets as in $\alpha$\,Cen.

\end{abstract}


\section{INTRODUCTION}

About 20\% of known extrasolar planets orbit one component of a stellar binary \citep{rag06,hag06}.  Both observation and 
theory indicate that the majority of these planets in binaries are similar to those orbiting single stars, owing to the wide ($a>
100$\,AU) stellar separations \citep{des07,hag06}.   However, there are a growing number of systems which present a major 
challenge to planet-formation modelers, because their {\em giant} planets orbit one component within closer binary systems.  
In order of discovery, these planetary systems are GJ\,86 (HR\,637, HD\,13445) \citep{que00}, $\gamma$\,Cephei \citep{hat03}, 
HD\,41004 \citep{zuc04}, HD\,196885 \citep{cor08}, and HD\,17605 \citep{mut10}.  All five of these binaries are thought to have
$a\la30$\,AU and a giant planet orbiting the primary star.  However, only in GJ\,86 is the stellar secondary a white dwarf rather 
than a second main-sequence star, and the Jovian planet thus orbits the originally {\em less} massive component.  Furthermore, 
the initial binary separation was smaller, making GJ\,86 a challenging environment in which to form planets.

Our view of planet formation in binaries has changed significantly in the past 17 years.  Simulations of binary stars with $a<40
$\,AU failed to retain sufficient circumstellar material to form planets around one star, and thus precluded planet birth by either the 
core accretion or disk-instability mechanisms (\citealt{nel00,art94}; see \citealt{pra10} for a detailed review).  Yet observers have 
imaged favorable planet-forming environments around the components of close binary stars, implying planet formation in these 
systems may be as common as around single stars \citep{ake98,mat94}.  For example, the two well-separated disks in the binary 
system L1551 \citep{rod98} indicate that, despite disk truncation, it is still possible for both components to retain a significant amount 
of their original circumstellar material ($0.03-0.06\,M_{\odot}$) in disks with considerable radii ($r\approx10$\,AU).  These disk masses 
are comparable to the minimum-mass model of the primordial Solar nebula \citep{hay81,wei77} suggesting that planet formation in 
these environments may proceed as in disks around single stars.

The Jovian planet\footnote{The planet is often referred to incorrectly as GJ\,86b, while the correct designation is GJ\,86Ab} orbiting 
GJ\,86A (K0\,V) was discovered via precision radial-velocity monitoring \citep{que00}, revealing a 15.8\,d orbit (0.11\,AU separation) 
with $m\sin{i}=4$\,$M_{\rm Jup}$.  At the time of discovery, an additional long-term radial velocity drift was reported, suggesting the 
presence of a more distant, unseen stellar companion. Subsequently, the distant companion was directly detected at a projected 
separation of $1\farcs7$, but 9\,mag fainter than the primary in the $K$ band \citep{els01}. The companion, GJ\,86B, was later 
imaged by \citet{mug05} and \citet{lag06} whose data clearly reveal orbital motion; between 2000 and 2005 the angular separation 
of the pair changed by $0\farcs25$ and the position angle changed by 17\degr.  GJ\,86B cannot be a low-mass star or brown dwarf: 
a substellar mass cannot be reconciled with the observed radial-velocity drift of GJ\,86A, and follow-up methane-band photometry 
showed GJ\,86B to have a near-IR color index near zero.  Moreover, $K$-band spectroscopy of GJ\,86B reveals an essentially 
Rayleigh-Jeans continuum source. Thus, GJ\,86B can only be a white dwarf \citep{lag06,mug05}.

\begin{table}
\begin{center}
\caption{{\em HST} Cycle 19 Observations of GJ\,86B\label{tbl1}} 
\begin{tabular}{@{}cccc@{}}

\hline

Instrument		&UT Date		&Filter/Grating	&Exposures\\		
				&			&			&(s)\\
\hline

WFC3 UVIS		&2012 Mar 31	&F225W		&$1.0\times16$\\
				&			&F225W		&$20.0\times4$\\
				&			&F275W		&$10.0\times4$\\
				&			&F336W		&$5.0\times4$\\
				&			&F390W		&$4.0\times4$\\
				&			&F438W		&$4.0\times4$\\
				&			&F555W		&$2.0\times4$\\
				&			&F625W		&$2.0\times4$\\
				&			&F814W		&$30.0\times4$\\

\hline			

STIS CCD			&2012 May 29	&G430L		&$100.0\times3$\\
				&			&G430M		&$600.0\times4$\\
				&			&G750M		&$400.0\times3$\\

\hline
\end{tabular}
\end{center}

\end{table}

In order to ideally constrain the formation and evolution of the giant planet orbiting GJ\,86A, it is necessary to obtain the 
orbit and component masses of the binary system.  The 3300:1 contrast ratio between components in the infrared makes this 
difficult but possible with adaptive optics, yet white dwarfs do not exhibit spectral features in this wavelength region.  Thus an 
optical study of GJ\,86B is necessary and a {\em Hubble Space Telescope (HST)} program was initiated to achieve this goal.  
In \S2 are described the imaging and spectroscopic observations, while \S3 describes the atmospheric modeling, derivation of
current (and progenitor) stellar parameters and binary orbit constraints.  New static limits on the GJ\,86 planetary system are 
calculated based on the {\em HST} observations and possible similarities with the Procyon system are discussed in \S4.

\section{OBSERVATIONS \& DATA REDUCTION}

Due to orbital motion observed between 2000 and 2005 \citep{lag06,mug05}, it was expected that precise spatial offsets might 
be necessary for proper placement of the faint companion within the spectroscopic aperture.  Thus, WFC3 imaging was executed 
for both photometric and astrometric purposes, prior to STIS spectroscopy.

\begin{table}
\begin{center}
\caption{Flux Densities \& Effective Wavelengths for GJ\,86B$^a$\label{tbl2}} 
\begin{tabular}{@{}cccc@{}}
\hline

Filter		&$\lambda_{\rm eff}^b$	&$F_{\lambda}\,/\,10^{-14}$			&$F_{\nu}\,/\,10^{-26}$\\
		&(\AA)				&(erg\,s$^{-1}$\,cm$^{-2}$\,\AA$^{-1}$)	&(erg\,s$^{-1}$\,cm$^{-2}$\,Hz$^{-1}$)\\

\hline

F225W	&2385				&3.49							&6.62\\
F275W	&2722				&4.04 							&9.99\\
F336W	&3355				&3.93 							&14.76\\
F390W	&3908				&3.40							&17.34\\
F438W	&4315				&3.02 							&18.75\\
F555W	&5237				&2.18 							&19.91\\
F625W	&6183				&1.56						 	&19.87\\
F814W	&7899				&0.80 							&16.74\\

\hline
\end{tabular}
\end{center}

$^a$ The flux of GJ\,86A was measured to be 31.4\,mJy in F225W.\\
$^b$ Effective wavelengths were derived by convolving the adopted stellar model in $F_{\lambda}$ with the WFC3 filter 
transmission curves.

\end{table}

\subsection{WFC3 Imaging}

\begin{figure*}
\includegraphics[width=172mm]{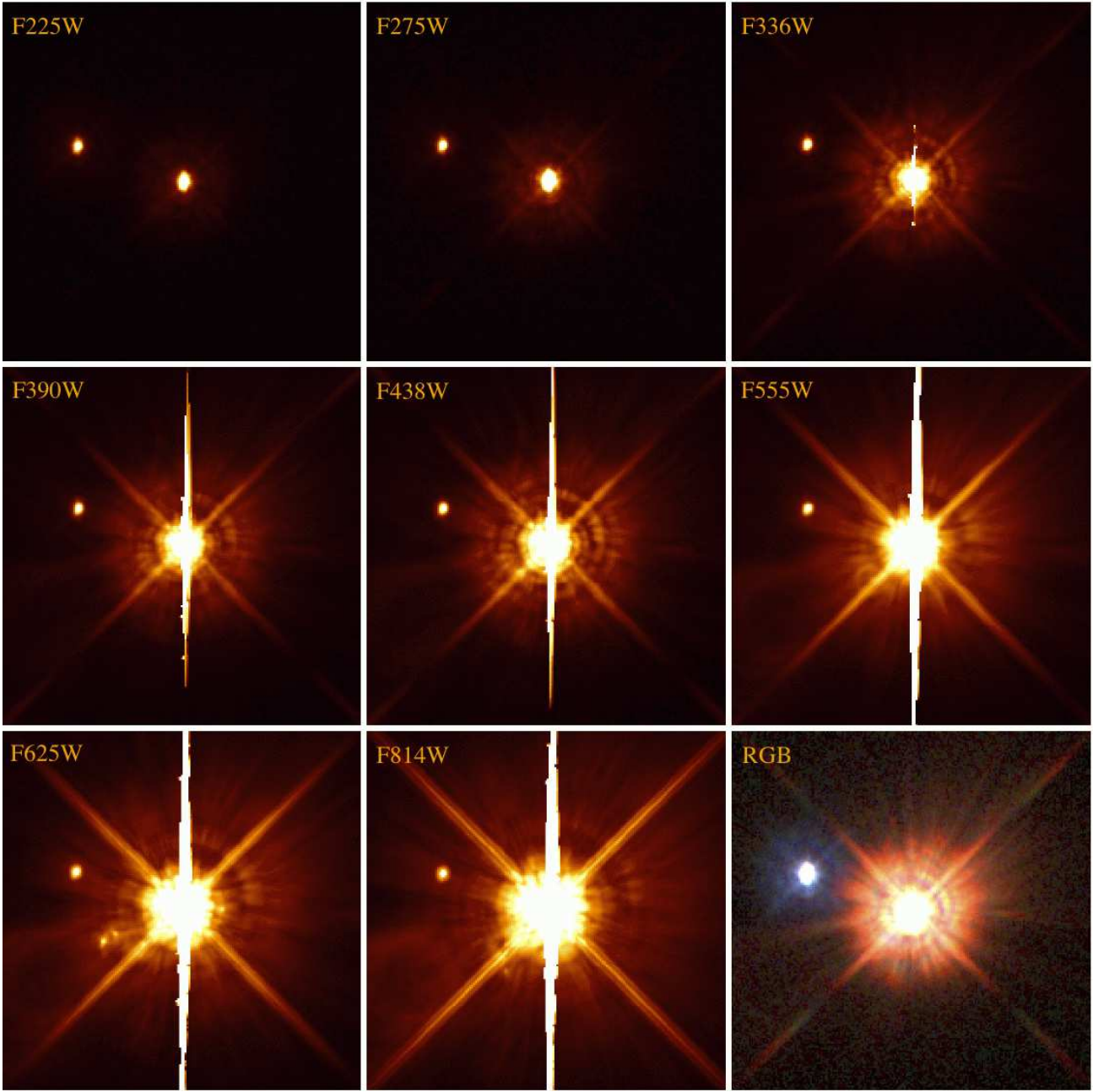}
\caption{Images of GJ\,86 in the eight specified ultraviolet and optical filters using WFC3 UVIS.  The images are oriented 
so that upward is 17\fdg60 east of north, and each frame is $8''$ on a side.  The bottom-right image is a composite RGB frame 
using the F225W (blue), F275W (green), and F336W (red) filters.  A series of short exposures at F225W were obtained with an 
unsaturated primary, and these data were used for the relative astrometry of the pair.
\label{fig1}}
\end{figure*}

GJ\,86 was imaged on 2012 March 31 using the Wide Field Camera 3 (WFC3) in the UVIS channel with a $512\times512$ 
(UVIS2-C512C-SUB) subarray.  The imaging sequence took place over a single orbit and began with a series of 1\,s exposures 
in the F225W filter.  In this short exposure time, the flux from the $V=6.2$\,mag, $T_{\rm eff}\approx5200$\,K  primary was expected 
to leave an unsaturated, linear response on the CCD.  This sequence employed the default UVIS dither pattern with four points in a parallelogram, and was repeated four times for a total of 16 exposures.  Both primary and secondary star were well-detected in this 
first set of exposures, and in the linear response regime with high signal-to-noise (S/N) $>500$.  This set of images was used to 
robustly measure the separation of the binary, where the secondary star was found at offset:

\begin{center}
\begin{tabular}{@{}ccc@{}}

$\alpha=2\farcs351(2)$	&$\theta=88\fdg96(4)$ (J2000)		&Date: 2012.2468\\

\end{tabular}
\end{center}

Immediately following the short exposure sequence, a single dither pattern of four deeper exposures was executed in each of 
eight filters (see Table \ref{tbl1} and Figure \ref{fig1}).  At the phase II design stage, the brightness of the companion was poorly 
constrained and the longer exposures were to insure high S/N photometry (up to the limit imposed by the PSF wings of the primary) 
at all wavelengths.  Fortunately, even without stray light correction, the dither-combined image sets yielded S/N $>500$ for GJ\,86B 
shortward of 5500\,\AA, and S/N $>200$ beyond.  Thus, the relatively faint companion was mostly unaffected by the primary.

\begin{figure*}
\includegraphics[width=172mm]{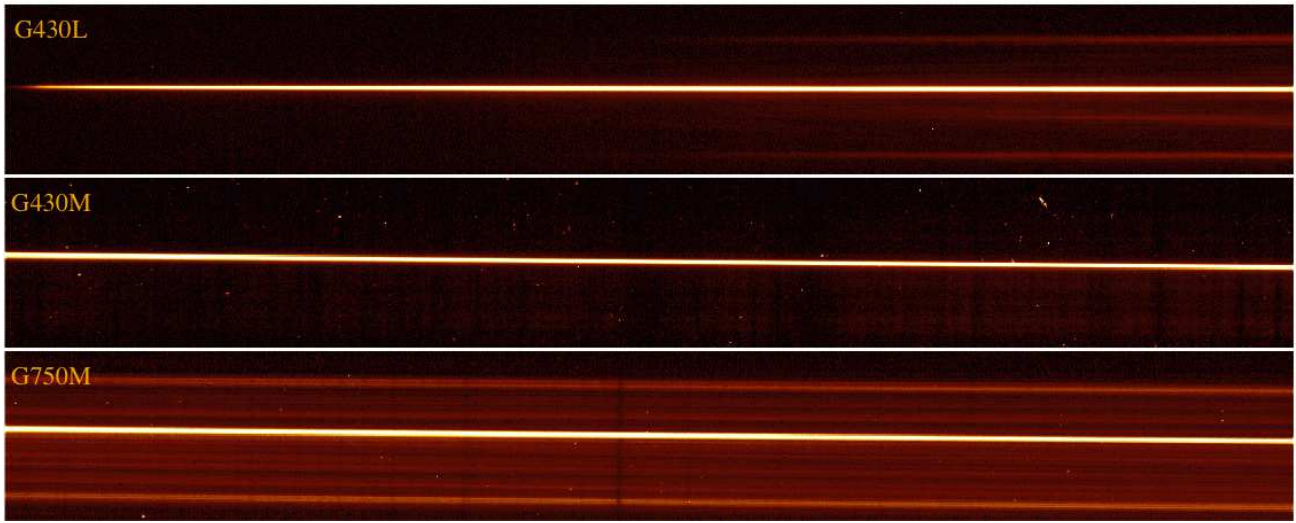}
\caption{Combined-frame STIS CCD spectral images of GJ\,86B taken in three gratings, with wavelength increasing to the right.  
The ambient light of the primary is relatively mild, even in red light.  Despite rejection routines that drastically reduced the number 
of cosmic rays, hot and cold pixels present in the single exposures, some artifacts remain in the combined 2D images, especially 
at the shortest wavelength setting.
\label{fig2}}
\end{figure*}

Photometry was performed on multi-drizzled images produced by the STScI software and data pipeline {\sc OPUS} 2012.1 
and {\sc CALWF3} 2.6.2.  The flux of GJ\,86B was measured in an $r=7$\,pixel aperture radius and corrected using the encircled 
energy data available in the WFC3 Instrument Handbook version 3.0 \citep{dre10}.  The aperture corrections were derived by 
interpolating the tabular values at the effective wavelength of the best fitting atmospheric model (see \S3) in each bandpass, 
and the process was iterated until there were no significant changes in the photometry or stellar model parameters; in practice 
this occurred after a single repetition.  The background flux of GJ\,86A was subtracted by examining points opposite, in $x$ and 
$y$ on the subarray, from the bright primary.  This ambient light correction resulted in flux measurement changes less than 1.2\% 
for the five filters below 5000\,\AA, but rose to around 4\% for the three longer wavelength filters.  The measured fluxes for GJ\,86B 
are listed in Table \ref{tbl2}.

\subsection{STIS Spectroscopy}

Observations with the Space Telescope Imaging Spectrograph (STIS) were executed on 2012 May 29, beginning with an acquisition 
sequence of GJ\,86A, followed by an offset and acquisition/peak-up of GJ\,86B based on the WFC3 astrometry.  Examination of the 
acquisition and peak-up images shows they were successful.  Spectroscopy was performed with the G430L, G430M, and G750M 
gratings as detailed in Table \ref{tbl1}, using default STIS dither patterns along the slit, and achieving average resolutions of 6.8, 
0.7, and 1.4\,\AA, respectively.

The individual spectral images were first shifted onto a common frame as indicated by the image header astrometry and using 
integer pixel shifts.  Next, the shifted images were combined into a single 2D frame using {\sc mscombine} in the IRAF STSDAS 
software package.  Prior to this step, the spectral images contained a significant number of bad pixels and cosmic rays, especially 
in the G430M observations.  Rejection of these unwanted pixels was performed using {\sc ccdclip} at $4\sigma$, although this 
could not completely remove the accumulated artifacts.  

Figure \ref{fig2} shows the final, combined spectral images for each of the grating settings.  Although stray light from the 
primary is apparent in all the images, and most readily seen in the red, the signal from GJ\,86B was essentially unaffected.  The 
contrast between the peak signals from the science target and the primary diffraction spikes was 17:1 at the lowest in the G750M 
observations, and at least twice as high in the region of the science spectrum.  

\begin{figure}
\includegraphics[width=86mm]{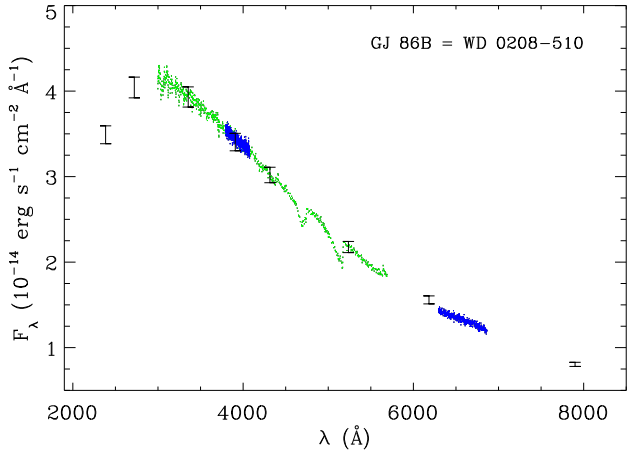}
\caption{Flux-calibrated WFC3 photometry and STIS spectra of GJ\,86B for each of the configurations specified in Table \ref{tbl1}.  
Figure \ref{fig4} plots the spectra in greater detail.
\label{fig3}}
\end{figure}

Lastly, 1D spectra were extracted using {\sc x1d}, and appropriate sky subtraction to correct for the relatively mild background.  
The flux-calibrated and normalized STIS spectra are displayed in Figures \ref{fig3} and \ref{fig4}.  The S/N per pixel was determined 
from continuum regions within each of the spectra to be close to 85, 88, and 53 in the G430L, G430M, and G750M gratings respectively.

\section{STELLAR \& BINARY PARAMETERS}

\begin{figure}
\includegraphics[width=86mm]{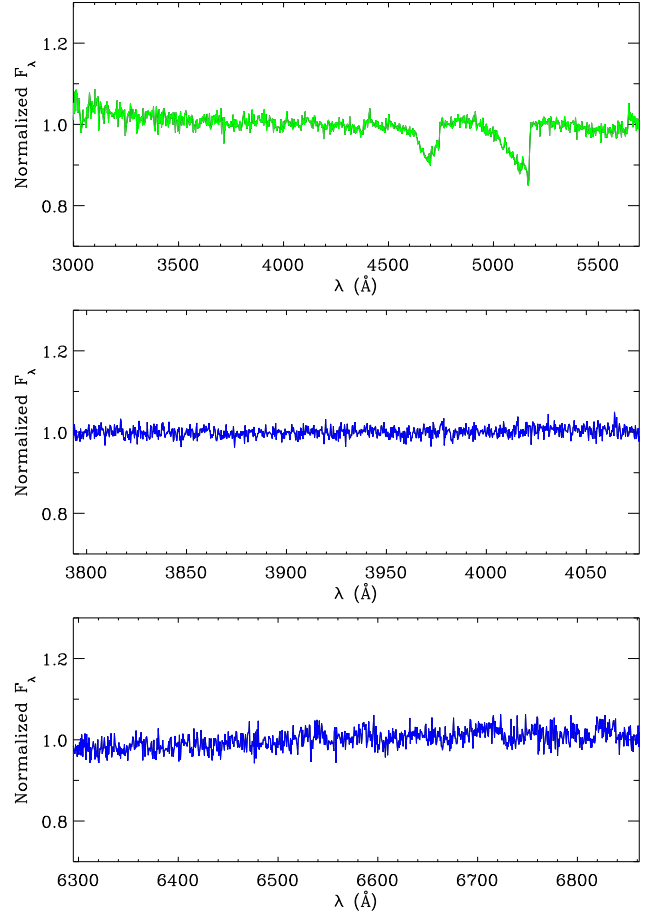}
\caption{From top to bottom are normalized STIS spectra of GJ\,86B obtained with the G430L, G430M, and G750M gratings.  
The only detected spectral features are the C$_2$ Swan bands.  Upper limits on calcium and hydrogen were derived from the 
Ca\,{\sc ii} K and H$\alpha$ regions (Table \ref{tbl3}).
\label{fig4}}
\end{figure}

The upper spectrum shown in Figure \ref{fig4} displays the distinctive and broad C$_2$ Swan bands, indicating that GJ\,86B has 
a DQ spectral classification.  This class of white dwarfs is characterized by helium-dominated atmospheres with trace carbon, likely 
dredged from the core \citep{pel86}.  Neither Ca\,{\sc ii} K nor H$\alpha$ were detected in the relatively deep and higher resolution 
G430M and G750M spectra.  Following convention the star has the white dwarf designation WD\,0208$-$510 \citep{mcc99}.

\subsection{Atmospheric Modeling}

Stellar parameters for GJ\,86B were derived in two ways, using model atmosphere grids appropriate for DQ white dwarfs (see 
\citealt{duf05} for details on the models and fitting techniques).  First, the effective temperature and solid angle were estimated from 
fitting the broad baseline WFC3 photometry.  These parameters were then combined with the Hipparcos parallax of 92.74\,mas 
\citep{van07,per97} to determine the stellar radius, which was then converted into mass using theoretical mass-radius relationships 
\citep{fon01}.  However, the shape of the energy distribution -- and thus the derived effective temperature -- is slightly sensitive to the 
amount of photospheric carbon.  Model fits to the C$_2$ Swan bands in the G430L STIS spectrum were used to measure the carbon 
abundance, with $T_{\rm eff}$ and $\log\,g$ fixed at the values obtained from the WFC3 photometry.  The effective temperature and 
solid angle were re-derived using C/He fixed to the spectroscopic value.  This procedure was iterated until all parameters converged, 
with a good fit obtained at $T_{\rm eff}=8420\pm90$\,K, $\log\,({\rm C/He})-4.58\pm0.15$.

\begin{table}
\begin{center}
\caption{Current \& Progenitor Stellar Parameters for GJ\,86AB\label{tbl3}} 
\begin{tabular}{@{}lr@{}}
\hline

\multicolumn{2}{c}{GJ 86B / WD\,0208$-$510}\\
\\

White Dwarf Remnant:				&\\
SpT								&DQ6\\
$V_{\rm eff}$ (mag)					&13.2\\
$T_{\rm eff}$ (K)					&$8180\pm120$\\
$\log\,g$ (cm\,s$^{-2}$)				&$8.02\pm0.02$\\
Mass ($M_{\odot}$)					&$0.59\pm0.01$\\
Radius ($R_{\odot}$)				&$0.01245\pm0.0015$\\
Cooling Age (Gyr)					&$1.25\pm0.05$\\
$\log\,({\rm C/He})$					&$-4.8\pm0.2$\\
$\log\,({\rm H/He})$					&$<-4.3$\\
$\log\,({\rm Ca/He})$					&$<-11.8$\\
\\

Main-Sequence Progenitor:			&\\
SpT								&A5\,V\\
Mass ($M_{\odot}$)					&$1.9\pm0.1$\\
Lifetime (Gyr)						&$1.4\pm0.2$\\

\hline

\multicolumn{2}{c}{GJ 86A / HR\,637 / HD\,13445}\\
\\

SpT								&K0\,V\\
$V$ (mag)						&6.1\\
$T_{\rm eff}$ (K)					&5200\\
Mass ($M_{\odot}$)					&0.80\\
Age (Gyr)							&2.5\\

\hline
\end{tabular}
\end{center}
{\em Note}.  Parameters listed for GJ\,86A are values representative of those found in the literature \citep{ghe10,van09,mam08,
val05,saf05,rib03,fly97}.
\end{table}

Second, because the STIS data are flux-calibrated with a high degree of confidence, the effective temperature and carbon abundance 
can be simultaneously obtained from the G430L spectrum alone, yielding $T_{\rm eff}= 8145\pm180$\,K, $\log\,({\rm C/He})=-4.82\pm
0.15$.  The difference in resulting model parameters is significant and most likely caused by the two shortest wavelength fluxes.  Indeed, 
the F225W and F275W bandpasses span a region influenced by at least one strong C\,{\sc i} line which models tend to over-predict (in
the core or wings).  An example of this was observed by \citet{pro02} in the STIS ultraviolet spectrum of Procyon\,B, where the strength of 
the 2478\,\AA line is much weaker (by nearly a factor of 30 in C/He) than predicted by models.  The model shortcomings are understood to 
be a consequence of the high density environments in cool helium-rich white dwarf photospheres; the classical Van der Waal broadening 
treatment within the impact approximation, traditionally used in DQ model atmospheres, is not appropriate for these lines \citep{koe82}. 

\begin{figure*}
\includegraphics[width=172mm]{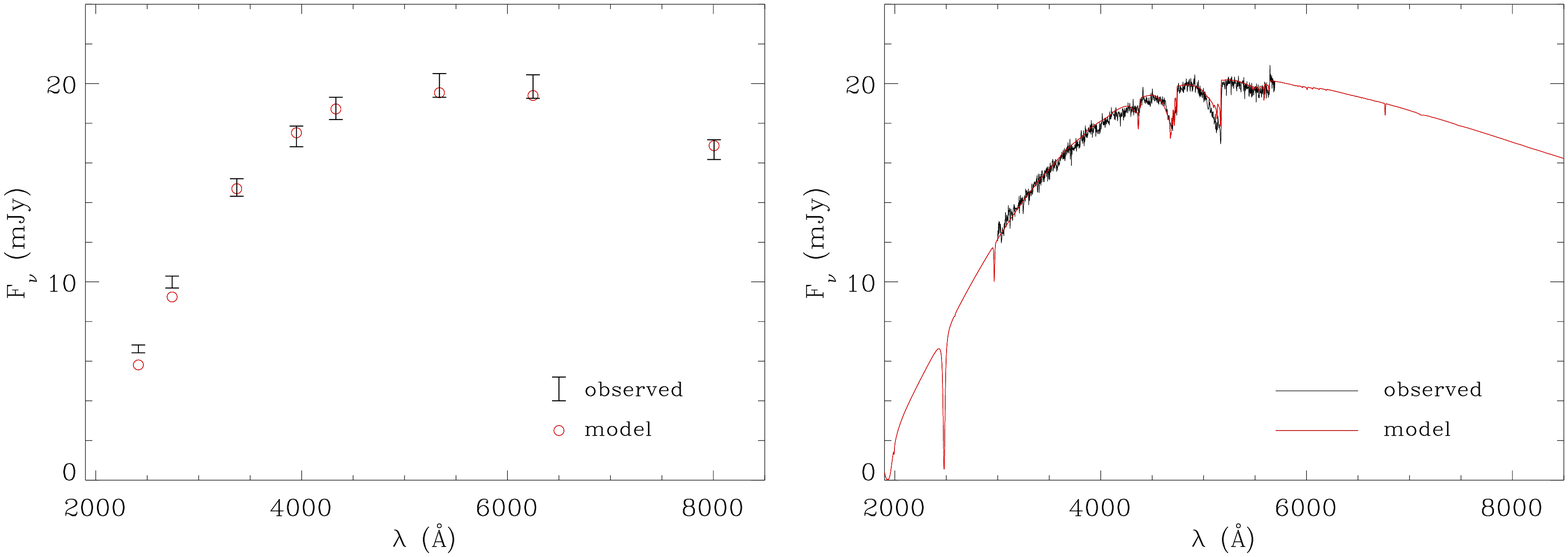}
\caption{Comparison between the observed and modeled stellar fluxes.  In the left-hand panel, the model fluxes agree well with the 
WFC3 photometry in all bandpasses except F225W and F275W (see \S3.1).  In the right-hand panel the model spectrum reproduces 
both the optical continuum and C$_2$ bands in the STIS G430L data, but indicates a strong C\,{\sc i} 2478\,\AA \ feature that under-predicts 
the stellar flux in the two ultraviolet filters.  All spectral features are due to atomic or molecular carbon.
\label{fig5}}
\end{figure*}

If one ignores the two ultraviolet fluxes affected by these strong C\,{\sc i} lines, and uses the remaining six WFC3 photometric data 
points, then $T_{\rm eff}=8210\pm180$\,K, $\log\,({\rm C/He})=-4.76\pm0.15$ is obtained, in agreement with the parameters from fitting 
the spectrum alone.  For all stellar and evolutionary parameters, the weighted average of the photometric (ignoring F225W and F275W) 
and spectroscopic effective temperatures, and the non-weighted average of the carbon abundances are thus adopted and listed in Table 
\ref{tbl3} ($T_{\rm eff}=8180\pm120$\,K and $\log\,({\rm C/He})=-4.8\pm0.2$).  Upper limit hydrogen and calcium abundances were also 
derived and are listed in Table \ref{tbl3}, while Figure \ref{fig5} shows the best fitting model compared to the photometry and G430L 
spectrum. All the derived stellar parameters are listed in Table \ref{tbl3}.

\subsection{Current Binary Orbit}

\begin{figure}
\includegraphics[width=84mm]{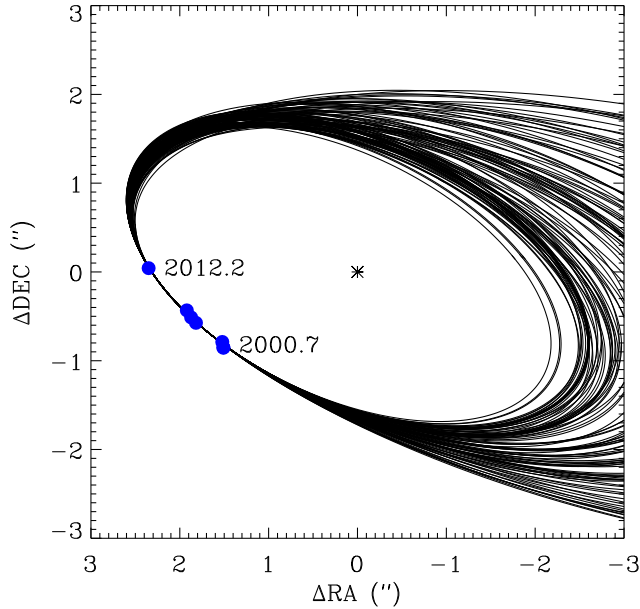}
\caption{The blue data points are the astrometric offsets determined here using WFC3 (leftmost point), together with those 
reported by \citet{lag06} and \citet{els01}.  An initial sample of 10\,000 orbital solutions were generated through a Monte Carlo 
search, and overplotted in black are the 100 orbits with the best $\chi^2$ values, within the restricted mass range $M_{tot}=1.4
\pm0.1\,M_{\odot}$.
\label{fig6}}
\end{figure}

The epoch 2012.2 WFC3 imaging provides a new point along the binary orbit of GJ\,86, and is shown in Figure \ref{fig6} together 
with previous astrometric measurements \citep{lag06,els01}.  Unfortunately, one quarter of the full orbit has not yet been observed, 
and hence the orbital parameters can only be loosely constrained.  For all orbital calculations, $M_A=0.8\,M_{\odot}$ and $M_B=
0.6\,M_{\odot}$ are adopted for simplicity; the actual component masses are not better constrained at present.

Using the new astrometric datum, a statistical analysis of orbits that fit the observations was performed.  A Monte Carlo simulation 
of 10\,000 orbital solutions was created with periods from 20 to 500\,yr, eccentricities from 0.00 to 0.99, and with time of periastron 
passage from 1800 to 2200.  From these solutions, the 100 orbits with the best $\chi^2$ values, and with total masses within the 
limited range $M_{tot}=1.4\pm0.1\,M_{\odot}$ were selected.  These orbits are plotted in Figure \ref{fig6}, and the resulting range 
in orbital parameters is listed in Table \ref{tbl4}. As can be seen, only the system inclination is significantly constrained at present.
The solutions with low eccentricities tend to have relatively short orbital periods, whereas the longer period solutions require high 
eccentricities.  Notably, the orbital solution with the shortest period yields a semimajor axis $a=27.8$\,AU with eccentricity $e=0.10$, 
and is significantly larger than the representative solution with $a=18.4\,$AU adopted by \citet{lag06}.

\section{BINARY \& PLANETARY EVOLUTION}

\subsection{Total System Age}

The white dwarf cooling age places a firm lower limit of 1.3\,Gyr on the lifetime of the system \citep{fon01}.  For the total age, one 
must add the hydrogen-burning lifetime of the progenitor, which should be in the range $1.2-1.6$\,Gyr for stars of mass $1.8-2.0\,
M_{\odot}$.  Thus, the evolved companion suggests a total system age of $2.65\pm0.25$\,Gyr.  Age estimates for the primary include 
$2.0-2.9$\,Gyr \citep{saf05} and 2.4\,Gyr \citep{mam08} based on activity-age relations, and these broadly agree with that derived for 
the secondary.  \citet{mam08} prefers an age of 3.7\,Gyr based on activity-rotation-age metrics, but this value is significantly older than 
expected for the total lifetime of the white dwarf and its progenitor.

\subsection{The Main-Sequence Progenitor System}

To constrain formation scenarios for the giant planet orbiting GJ\,86A, the binary orbit and component masses when both stars 
were on the main sequence is needed.  Using initial-to-final mass relations with some empirical constraints at the low-mass end 
of progenitor and remnant masses \citep{wil09,kal08}, GJ\,86B likely descended from a main-sequence, A-type star with a mass 
near 1.9\,$M_{\odot}$.  With both stars on the main sequence, the semimajor axis of the binary was smaller by a factor \citep{jea24} 

\begin{equation}
\frac{a_0}{a} =  \frac{M_B + M_A}{M_{B_0} + M_A} 
\label{eqn1}
\end{equation}

\noindent
This ratio is not very sensitive to the range of allowed masses, and yields  $a_0=0.52a$ for $M_{A}=0.80\,M_{\odot}$, $M_{B}=
0.59\,M_{\odot}$, $M_{B_0}=1.86\,M_{\odot}$.  A canonical mass value has been assumed for the primary, but is consistent with 
isochrones and various literature estimates \citep{van09,val05,rib03,fly97}.  From the current binary orbit simulations, a lower 
limit on the initial semimajor axis is obtained $a_0\geq14.5$\,AU.  Even at this lower limit, the 0.11\,AU planetary orbit of GJ\,86Ab 
is stable for binary eccentricities $e<0.9$ \citep{hol99,rab88}.  

\subsection{Planetesimals in GJ\,86}

Although the class of detached binary stars with a white dwarf and a main-sequence component of spectral type K or earlier are 
referred to as Sirius-type binaries \citep{hol08}, GJ\,86 is also notably {\em Procyon-like}.  Table \ref{tbl5} lists the atmospheric and 
stellar parameters for these two white dwarf companion systems, demonstrating their broad similarities.  

Because GJ\,86 is known to host a planetary system, with at least one giant planet orbiting the primary, it may also support a 
planetesimal population.  These surviving minor bodies could be gravitationally perturbed into close encounters with the white 
dwarf, where they would be tidally shredded and eventually accreted, polluting the atmosphere with heavy elements \citep{jur03}, 
as observed at nearly 30 single white dwarfs \citep{gan12,far12,kil12,deb12}.  In a binary such as GJ\,86, stable planetesimal belts 
can persist as circumstellar rings at either star, or a circumbinary disk.  

\begin{table}
\begin{center}
\caption{Orbital Constraints from Monte Carlo Simulations\label{tbl4}} 
\begin{tabular}{@{}cl@{}}
\hline

Parameter			&Range\\

\hline

$P$					&$120-481$\,yr\\
$T_0$				&$1933-2067$\\
$e$					&$ 0.00-0.61$\\
$a$					&$2\farcs57-6\farcs47$\\
					&$27.8-69.8$\,AU\\
$i$ 					&$114\fdg7-122\fdg6$\\
$\Omega$ 			&$63\fdg7-76\fdg1$\\
$\omega$	 			&$0\degr-358\degr$\\ 

\hline
\end{tabular}
\end{center}
{\em Note.}  The upper range of periods and semimajor axes result from the simulation input $P_{max}=500$\,yr, but these 
orbits require high eccentricities $e>0.5$ and are thus less likely.
\end{table}

One region that remains stable over both the current and former binary configurations and $e<0.1$ is within roughly 3\,AU of the 
K star primary.  Stable circumstellar regions for the white dwarf are complicated by competing processes: while the region within 
2\,AU will be depleted of small bodies by direct engulfment and tides during the giant phases \citep{mus12}, surviving objects
will have their orbit expanded due to mass loss, by a factor of $M_{B_0}/M_B\approx3.2$.  However, the critical radius for stability 
also increases due to mass loss, but by a factor smaller than $({M_B + M_A})/({M_{B_0} + M_A})\approx1.9$, and therefore a narrow 
range of stable, planetesimal orbits are possible for the white dwarf if $a\ga30$\,AU currently.  It is worth noting that these calculations 
are purely static, and thus ignore the dynamic effects of mass loss which may have significant impact on stable orbits \citep{ver12,ver11}.  
Circumbinary orbits beyond 80\,AU are also stable in the current and former binary configurations for $a\approx30$\,AU and $e<0.1$.

Thus, one of the science goals of the {\em HST} observing program was to search for evidence of planetesimals in this system, via 
atmospheric metal pollution in the white dwarf.  As discussed in \S2.2, calcium absorption was not detected in the STIS data, and a 
strict upper limit of $\log\,({\rm Ca/He})<-11.8$ was derived for GJ\,86B.  This corresponds to $M_{\rm Ca}<3.2\times10^{17}$\,g in 
the stellar convection zone, or an upper limit to the total accreted heavy element mass of $M_{\rm Z}\la2.0\times10^{19}$\,g for 
material that is 1.6\% calcium by mass as in the bulk Earth \citep{all95}.  Because the Ca\,{\sc ii} K line is the most sensitive tracer 
of metal pollution at optical wavelengths for white dwarfs of this effective temperature \citep{koe05,zuc03}, this large cometary-sized
mass is the strongest possible limit to the observable planetesimal population.

\subsection{A Failed Planetary System at Procyon?}

\begin{table}
\begin{center}
\caption{Comparison of GJ\,86B \& Procyon\,B \label{tbl5}} 
\begin{tabular}{@{}lrr@{}}
\hline

Parameter				&GJ\,86B				&Procyon\,B$^a$\\

\hline

SpT						&DQ6.2				&DQZ6.5\\
$T_{\rm eff}$ (K)			&8180				&7740\\
Mass ($M_{\odot}$)			&0.59				&0.60\\
$\log\,({\rm H/He})$			&$<-4.3$				&$<-3.7$\\
$\log\,({\rm C/He})$			&$-4.8$				&$-5.5$\\
$\log\,({\rm Mg/He})^b$		&...					&$-10.4$\\
$\log\,({\rm Ca/He})$			&$<-11.8$			&$-11.8$\\

\hline

Primary SpT				&K0\,V				&F5\,IV\\
$a$ (AU)					&$\geq28$			&15.2\\
$e$						&...					&0.41\\

\hline

\end{tabular}
\end{center}
$^a$ Atmospheric parameters from \citet{pro02}.\\
$^b$ The STIS spectra of GJ\,86B do not cover the Mg\,{\sc ii} 2800\,\AA \ resonance lines.
\end{table}

Interestingly, and also using STIS, \citet{pro02} found that Procyon\,B exhibits a strong Mg\,{\sc ii} resonance line at 2800\,\AA,
a few Fe lines in the near-ultraviolet, and Ca\,{\sc ii} H and K absorption in the optical.  While Procyon is not known to host any 
planets or planetesimal debris (via precision radial velocity, transits, astrometry, direct imaging, or infrared excess), the white 
dwarf is externally polluted with at least three heavy elements that could have originated in a minor planetary body \citep{far10}.  
Using the published magnesium abundance, Procyon\,B has accreted at least $M_{\rm Mg}=5.1\times10^{18}$\,g within the past 
few $10^7$\,yr, where 2.8\,Myr is a single sinking timescale for this element at the stellar effective temperature \citep{koe09}.  For
material with a chondritic or Earth-like elemental abundance, this is equivalent to a total accreted heavy element mass of at least 
$3-6\times10^{19}$\,g \citep{lod98,all95}, and equivalent to a $25-35$\,km diameter object for typical asteroid densities.

Other possible sources for the atmospheric metals are the interstellar medium, or the stellar wind from Procyon\,A \citep{pro02}.
In both of these cases the accreted matter would have solar abundances ($\log\,({\rm Mg/H})=-4.45$, $\log\,({\rm Mg/He})=-3.35$;
\citealt{lod03}) and imply time-averaged accretion rates of $\dot M_{\rm H}=8.0\times10^{8}$\,g\,s$^{-1}$ and $\dot M_{\rm He}=2.6
\times10^{8}$\,g\,s$^{-1}$.  If the heavy elements have been continually captured from Procyon\,A, then hydrogen and helium should
have accumulated in the white dwarf atmosphere over its 1.4\,Gyr cooling age \citep{fon01}.  During this period $3.6\times10^{25}$\,g 
of hydrogen and $1.1\times10^{25}$\,g of helium will have been added to the $2.1\times10^{28}$\,g helium convection zone, yielding
a current abundance $\log\,({\rm H/He})=-2.2$.  However, the upper limit derived from the lack of H$\alpha$ absorption is $\log\,({\rm 
H/He})<-3.7$ and inconsistent with stellar wind accretion.  

If one ignores the issue of hydrogen accumulation, the mass loss rate from Procyon\,A that is necessary to supply the magnesium 
in the white dwarf at the observed time-averaged rate is $1.8\times10^{-9}\,M_{\odot}$\,yr$^{-1}$ for a Bondi-Hoyle accretion flow 
\citep{deb06} flow and $9.1\times10^{-8}\,M_{\odot}$\,yr$^{-1}$ for gravitational (Eddington) accretion.  These mass-loss rates are 
between 5 and 7 {\em orders of magnitude} larger than that measured for the Sun and $\alpha$\,Cen \citep{woo01,wit89}.  Therefore, 
wind capture can be ruled out with confidence.  Similar arguments can be used to discount the accretion of interstellar matter.  First, 
hydrogen accumulates whereas metals sink, and thus the hydrogen-deficient atmosphere of Procyon\,B is at odds with the accretion 
of solar abundance material for more than 2\% of its cooling lifetime.  Second, molecular cloud densities are required to account for 
the observed metals \citep{far10}, but Procyon has been moving within the local, $r\sim100$\,pc, ISM-poor Bubble \citep{red08} for 
at least 2.6\,Myr.

The remaining possibility is the accretion of one or more planetesimals.  Accounting for both the current and former binary 
configurations, stable orbital regions at Procyon can be either circumbinary beyond 55.4\,AU or circumprimary within 2.3\,AU 
\citep{hol99}, although bodies in the latter region would be dynamically unlikely to be perturbed towards the surface of the white 
dwarf.  There is no circumstellar region at Procyon\,B that remains stable over the lifetime of the binary.  Thus, a circumbinary 
population of planetesimals is the most likely source of the metals observed in Procyon\,B.  Observations with {\em Herschel} 
PACS at 160\,$\mu$m have not detected a far-infrared excess at Procyon, and limit the fractional dust luminosity to $L/L_*<5\times
10^{-7}$ over the $30-300$\,AU region (G. Kennedy 2012, private communication).  This limit is only a few times greater than the
fractional dust luminosity of the Kuiper belt \citep{boo09,ste96}, and for similar particles would result in an upper dust mass limit
of roughly an Earth mass for Procyon.

On the main sequence, Procyon should have had a semimajor axis near 9\,AU.  Even for $e=0.1$ this implies that disk material
available for planet building would be confined to within 2.0\,AU of Procyon\,A and 2.4\,AU of Procyon\,B, and thus completely
within their respective snow lines.  Thus, giant planet formation at either component of the Procyon system is likely precluded, and 
likely also for the $a>55$\,AU stable circumbinary environment.  Only the inner regions of Procyon\,A appear capable of forming 
small, solid planets (as in $\alpha$\,Cen\,AB; \citealt{dum12}), that persist to the present day, assuming the binary eccentricity was 
not larger when both stars were on the main sequence.  Thus, the Procyon system may represent a case of failed or truncated planet 
formation, where large planetesimals were formed but further growth was prohibited.

\section{SUMMARY \& OUTLOOK}

High-contrast optical imaging and low-resolution spectroscopy with {\em HST} have unambiguously characterized the stellar 
companion to the planet-host star GJ\,86A.  The secondary is a relatively cool white dwarf with a helium-dominated atmosphere 
and molecular absorption bands due to trace carbon, which is almost certainly dredged from the core.  Neither H$\alpha$ nor 
Ca\,{\sc ii} K are detected in deeper, medium-resolution spectroscopy, the latter placing modest limits on the presence of
scattered planetesimal material.

The binary separation of GJ\,86 has continued to increase since last observed in 2005, and the actual orbit deviates from prior 
estimates.  While one-quarter orbit has not yet been observed, Monte Carlo simulations constrain the semimajor axis to $a\geq
27.8$\,AU, and the system inclination within 114\fdg7 $\leq i \leq122\fdg6$.  For planetary-binary coplanar orbits, these results 
imply the mass of GJ\,86Ab lies between 4.4 and 4.7\,$M_{\rm Jup}$.

Future imaging is necessary to observe apastron and dynamically constrain the component masses.  This can be done with adaptive 
optics in the near-infrared, but superior data will come from follow up imaging and spectroscopy with {\em HST} in the near-ultraviolet.  
In this relatively low-contrast wavelength regime, there will be narrow absorption lines from C\,{\sc i}, and possibly Mg\,{\sc ii} which is
a strong absorption feature in Procyon\,B.  Such spectral lines can provide a baseline radial velocity and gravitational redshift for the 
secondary; these would more efficiently characterize the current and former binary configurations than near-infrared data.

\section*{ACKNOWLEDGMENTS}

The authors thank an anonymous referee for a timely and thorough report.  J. Farihi thanks J. Provencal for helpful feedback on the 
available spectra of Procyon\,B, as well as G. Kennedy and M. Wyatt for a useful summary of the infrared excess and dust limits in this 
system.  This work is based on observations made with the {\em Hubble Space Telescope} which is operated by the Association of 
Universities for Research in Astronomy under NASA contract NAS 5-26555.  These observations are associated with program 12548.  
Support for Program number 102548 was provided by NASA through grant HST-GO-12548 from the Space Telescope Science Institute.  
J. Farihi gratefully acknowledges the support of the STFC via an Ernest Rutherford Fellowship and as a PDRA.  N. Haghighipour 
acknowledges support from NASA grants EXOB NNX09AN05G and HST-GO-12548.06-A.  J. Holberg acknowledges support from 
NSF Grant AST-1008845.  M. Barstow and M. Burleigh acknowledge support from the STFC.

\label{lastpage}

\end{document}